\input{aipcheck}

\documentclass[
    ,final            
  ]
  {aipproc}

\layoutstyle{6x9}

\begin{document}

\title{Solar analogues in open clusters: The case of M67}

\classification{98.20.Di; 97.10.Ri; 97.10.Vm; 97.10.Wn}
\keywords      {Open clusters and associations: individual: M67 -- Stars: fundamental 
parameters -- Stars: late-type}

\author{K. \,Biazzo}{
  address={INAF - Catania Astrophysical Observatory, Catania, Italy}
 ,altaddress={ESO - European Southern Observatory, Garching bei M\"unchen, Germany}
}

\author{L. \,Pasquini}{
  address={ESO - European Southern Observatory, Garching bei M\"unchen, Germany}
}

\author{P. \,Bonifacio}{
  address={GEPI, Observatoire de Paris, Meudon, France}
  ,altaddress={INAF - Trieste Astronomical Observatory, Trieste, Italy} 
}

\author{S. \,Randich}{
  address={INAF - Arcetri Astrophysical Observatory, Arcetri, Italy}
}

\author{L. R. \,Bedin}{
  address={Space Telescope Science Institute, Baltimore, United States}
}

\begin{abstract}
Solar analogues are fundamental targets for a better understanding of our Sun and 
our Solar System. Usually, this research is limited to field stars, which offer 
several advantages and limitations. In this work, we present the results of a 
research of solar twins performed for the first time in a open cluster, namely M67. 
Our analysis allowed us to find five solar twins and also to derive a solar colour of 
$(B-V)_\odot=0.649\pm0.016$ and a cluster 
distance modulus of $9.63\pm0.08$. This study encourages us to apply the same method 
to other open clusters, and to do further investigations for planet search in the solar 
twins we find.
\end{abstract}

\maketitle


\section{Introduction}
The search for solar analogues and for exo-planets has been mostly focused around field stars. Such stars 
present several advantages, for instance a wide range of stellar characteristics 
(mass, age, effective temperature, chemical composition, etc.), which allows us to 
study the dependence of planet formation on stellar parameters. However, any astrophysical 
parameters cannot be constrained {\it a priori} in field stars. This was the fundamental 
reason leading us to begin for the first time in a open cluster the research for solar analogues. 
In fact, they show homogeneous age and chemical composition, common birth and early dynamical environment 
(\cite{Randich2005}), providing us an excellent laboratory for investigating the physics of planetary system 
formation. The old open cluster M67 is to this purpose a perfect target, having many 
solar-type stars and showing an age encompassing that of the Sun ($3.5-4.8$ Gyrs; 
\cite{Yadav2008}), a solar metallicity ([Fe/H]=$0.03\pm0.02$, \cite{Randich2006}) and lithium depleted 
G stars (\cite{Pasquini1997}). 

The present paper is the culmination of a work involving the chemical determination of M67 
(\cite{Randich2006}), photometry and astrometry (\cite{Yadav2008}) to obtain membership, and 
FLAMES/GIRAFFE high resolution spectroscopy to clean this sample from binaries, and to look for the best solar 
analogues using the line-depth ratios method and the wings of the H$\alpha$ line for deriving 
accurate effective temperatures with respect to the Sun ($\Delta T^{\rm LDR}$ and $\Delta T^{\rm H\alpha}$; \cite{Pasquini2008}).

\section{Sample selection and Observations}
We acquired spectra of M67 for 2.5 hours during three nights in February 2007 with the 
multi-object FLAMES/GIRAFFE spectrograph at the UT2/Kueyen ESO-VLT in Paranal 
(Chile). We chose the HR15N MEDUSA mode, which allows us to cover the spectral 
range 6470-6790 \AA, catching the H$\alpha$ and the lithium lines in one shot. 
With this configuration, the resolution of 17\,000 gives us the possibility to obtain 
for almost 100 stars good radial velocities and to determine effective temperature and 
lithium abundance. We have chosen from the catalogue of \cite{Yadav2008} bright stars ($V=13.0-15.0$ mag) 
with $(B-V)$ close to that of the Sun ($0.60-0.75$) and showing the best combination of proper motions measurements 
($\mu_{\alpha}\cos \delta$, $\mu_{\delta}$) and 
proper-motion membership probability ($P_{\mu}$) allowed us to observe at a time almost 100 
stars (\cite{Pasquini2008}). 

\section{Results}

\subsection{Radial Velocity}
\begin{figure*}[ht]
\includegraphics[width=7.5cm]{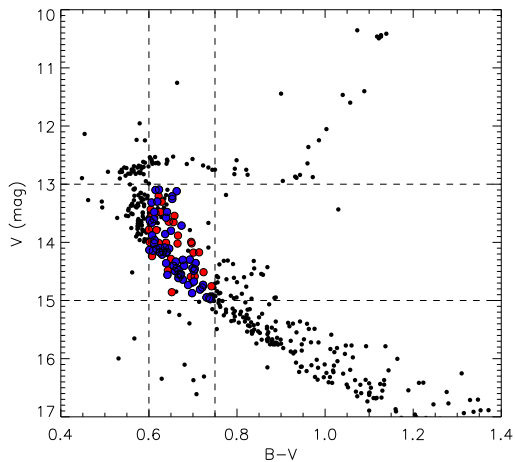}
\caption{\footnotesize M67 colour-magnitude diagram (\cite{Yadav2008}). In blue the 59 retained 
single member candidates, while in red the stars discarded are shown.}
\label{fig:cmd}
\end{figure*}

From the radial velocity variations of our stars observed during our run, 
we find that 59 of them are probable single cluster members with an average radial velocity 
$<V_{\rm rad}>=32.9\pm0.73$ km s$^{-1}$ (Fig.~\ref{fig:cmd}; \cite{Pasquini2008}). 

\subsection{Effective Temperature}
\subsubsection{Line-depth ratios}
In the spectral range covered by FLAMES/GIRAFFE, we have selected six line pairs sensitive to effective 
temperature and applied a method based on line-depth ratios (LDRs; \cite{GrayJoha1991,Catalano2002,Biazzo2007a,Biazzo2007b}). 
Thus, we have developed appropriate LDR$-T_{\rm eff}$ calibrations on synthetic spectra computed 
in the $T_{\rm eff}$ range 5400--6300 K and derived the effective temperature of our probable single members. 

\subsubsection{H$\alpha$ wings}
\begin{figure*}[ht]
\includegraphics[width=10cm]{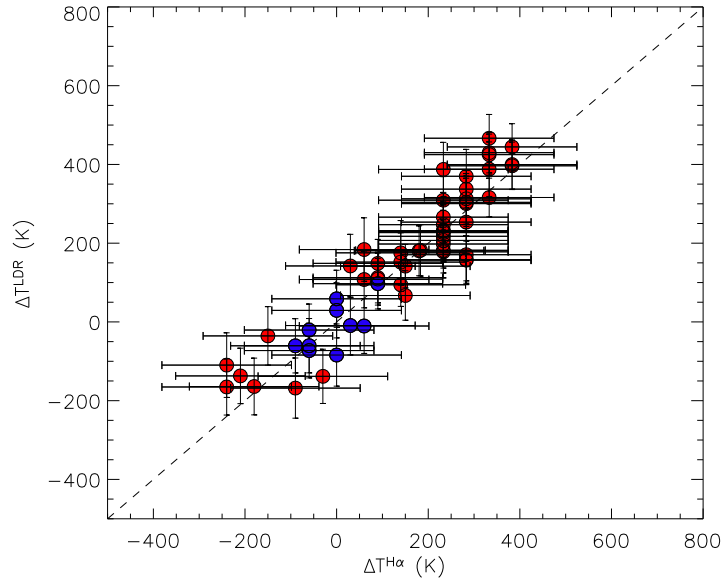}
\caption{\footnotesize $\Delta T^{\rm LDR}$ versus $\Delta T^{\rm H\alpha}$ for the probable single members. 
In blue the ten solar analogues are displayed.}
\label{fig:deltaT}
\end{figure*}

Since the wings of the H$\alpha$ line profile are very sensitive to temperature, we have also studied the 
behavior of this diagnostics. According to \cite{Cayrel1985}, the effective temperature 
of a star can be derived from the strength of its H$\alpha$ wings measured between 3 and 5 \AA~from 
the H$\alpha$ line-center, as compared to synthetic spectra H$\alpha$ line-wings in the same wavelength 
interval. In Fig.~\ref{fig:deltaT} we compare the results we obtain with the LDR and H$\alpha$-wing methods.

\subsection{Lithium Abundance}
Lithium is an element easily destroyed in the stellar interiors. It is important for understanding the 
complex interactions taking place in the past between the stellar external layers and the hotter interior. 
We have derived its abundance finding that many main sequence stars share the same lithium abundance of the Sun, indicating 
a similar mixing history. Furthermore, for the first time, we see the Li extra-depletion appears 
in stars cooler than 6000 K (\cite{Pasquini2008}).

\section{Solar analogues}
Comparing the $\Delta T^{\rm LDR}$ and the $\Delta T^{\rm H\alpha}$ and taking into account our lithium 
abundance determination, we find ten solar analogues (Fig.~\ref{fig:deltaT}). In particular, five stars are the 
closest to the Sun, with effective temperature derived with both methods within 60 K from the solar one. 

\subsection{$B-V$ solar colour}
Taking into accunt a reddening towards M67 of $E(B-V)=0.041\pm0.004$ (\cite{Taylor2007}), we have obtained a 
solar $(B-V)$ of $0.649\pm0.016$, which is in the middle between the old determinations (around 0.66--0.67) 
and the recent estimations (around 0.62--0.64). 

\subsection{Cluster distance modulus}
At the same way we computed the solar $(B-V)$, we have derived the cluster distance. A value of $9.63\pm0.08$ 
was found, in excellent agreement with the most recent determinations (\cite{Pasquini2008}). 

\section{Conclusions}
By using spectroscopic observations performed with FLAMES/GIRAFFE, we have found the best 
solar twins in M67, solar colour, and cluster distance modulus. Thanks to our promising results, we plan 
from one side to observe our single candidates at very high resolution 
for exo-planet research and from the other side to apply our method to other open clusters younger and 
older than the Sun in order to study the ``Sun in time''.








\begin{thebibliography}{100}

\bibitem{Biazzo2007a}
Biazzo, K., et al., \emph{Astronomische Nachrichten} \textbf{328}, 938 (2007a)

\bibitem{Biazzo2007b}
Biazzo, K., et al., \emph{Astronomy and Astrophysics} \textbf{475}, 981 (2007b)

\bibitem{Catalano2002}
Catalano, S., et al., \emph{Astronomy and Astrophysics} \textbf{394}, 1009 (2002)

\bibitem{Cayrel1985}                       
Cayrel, R., et al., \emph{Astronomy and Astrophysics} \textbf{146}, 249 (1985)

\bibitem{GrayJoha1991}			
Gray, D. F., \& Johanson, H. L., \emph{PASP} \textbf{103}, 439 (1991)

\bibitem{Pasquini1997}			
Pasquini, L., et al., \emph{Astronomy and Astrophysics} \textbf{325}, 535 (1997)

\bibitem{Pasquini2008}			
Pasquini, L., et al., \emph{Astronomy and Astrophysics} \textbf{489}, 677 (2008)

\bibitem{Randich2005}
Randich, S., et al., \emph{Msngr} \textbf{121}, 18 (2005)

\bibitem{Randich2006} 			
Randich, S., et al., \emph{Astronomy and Astrophysics} \textbf{450}, 557 (2006)

\bibitem{Taylor2007}
Taylor, B. J., \emph{Astronomical Journal} \textbf{133}, 370 (2007)
\bibitem{Yadav2008} 			
Yadav, R. K. S., et al., \emph{Astronomy and Astrophysics} \textbf{484}, 609 (2008)

\end{thebibliography}



\end{document}